\documentclass[%
 reprint,
superscriptaddress,
nofootinbib,
 amsmath,amssymb,
 aps,
 prd,
showkeys
]{revtex4-2}

\usepackage{graphicx}
\usepackage{dcolumn}
\usepackage{bm}

\usepackage{multirow}%
\usepackage{amsmath,amssymb,amsfonts}%
\usepackage{amsthm}%
\usepackage{mathrsfs}%
\usepackage[title]{appendix}%
\usepackage{xcolor}%
\usepackage{textcomp}%
\usepackage{booktabs}%
\usepackage{algorithmicx}%
\usepackage{algpseudocode}%
\usepackage{listings}%

\usepackage{color} 

\begin{document}

\preprint{APS/123-QED}

\title{Empirical Reconstruction of the JSNS$^2$ KDAR $\nu_\mu$-$^{12}$C Missing-Energy Spectrum with a Two-Ex-Gaussian and Generalized-Tail Model}

\author{Kyung Kwang Joo}
\email{kkjoo@chonnam.ac.kr}
\affiliation{Center for Precision Neutrino Research (CPNR), Department of Physics, Chonnam National University, Yongbong-ro 77, Puk-gu, Gwangju 61186, Republic of Korea}

\author{Jubin Park}
\email[Corresponding author: ]{honolov@ssu.ac.kr}
\affiliation{Department of Physics and Origin of Matter and Evolution of Galaxy (OMEG) Institute, Soongsil University, Seoul 06978, Republic of Korea}

\author{Minkyu Lee}
\email{rkftor28gh@soongsil.ac.kr}
\affiliation{Department of Physics and Origin of Matter and Evolution of Galaxy (OMEG) Institute, Soongsil University, Seoul 06978, Republic of Korea}

\author{Myung-Ki Cheoun}
\email{cheoun@ssu.ac.kr}
\affiliation{Department of Physics and Origin of Matter and Evolution of Galaxy (OMEG) Institute, Soongsil University, Seoul 06978, Republic of Korea}


\date{\today}

\begin{abstract}
Recent analyses of the JSNS$^2$ monoenergetic $\nu_\mu$ scattering on $^{12}$C at 235.5~MeV have compared 
the measured missing-energy spectrum with several nuclear models, including \textsc{NuWro}, \textsc{GiBUU}, and RMF+Achilles. 
While these models reproduce the overall peak position, their respective 
$\chi^2$ values of $35.5$, $176.8$, and $58.1$ 
indicate that none can simultaneously describe the spectral width and the high-energy tail, 
reflecting limitations in the treatment of binding energy, two-particle--two-hole (2p--2h) excitations, 
and final-state interactions (FSI). 
To address these discrepancies, we introduce an empirical yet physically motivated representation of the spectrum 
based on two exponentially modified Gaussian (ex-Gaussian) components for p- and s-shell knockout 
and a generalized power--exponential continuum term describing multinucleon and FSI-induced strength. 
The fit reproduces the JSNS$^2$ data within the fitted energy range with 
$\chi^2=8.0$ for 6 degrees of freedom. 
yielding parameters that quantify asymmetric broadening of the s-shell while preserving a narrow quasielastic p-shell response. 
This compact model demonstrates that a minimal empirical framework can capture key features of the nuclear response 
and provides a useful reference for phenomenological comparisons and future studies of quasielastic 
and 2p--2h dynamics in the few-hundred-MeV regime.
\end{abstract}

\maketitle


\section{Introduction}\label{sec1}
Neutrino--nucleus interactions in the few-hundred-MeV regime provide a uniquely sensitive arena in which quasielastic (QE) scattering, multinucleon excitations, and inelastic channels coexist and interfere, with observable consequences set by nuclear binding energy and final-state interactions (FSI). 
A quantitatively reliable description of these effects is indispensable both for oscillation-era neutrino physics and for isolating few-body correlations within complex nuclei\,\cite{Mahn:2018mai}.
Monoenergetic neutrino sources are especially valuable because they suppress flux uncertainties and turn final-state observables into direct probes of the nuclear response to the weak-current\,\cite{Mahn:2018mai,Volpe2000}.
In parallel, JSNS$^2$ has independently characterized the electron-neutrino flux 
using the $^{12}$C$(\nu_e, e^-)^{12}$ $\mathrm{N}_{g.s}$ reaction, 
providing complementary constraints on the source and detector response \cite{JSNS2:2024uxo}.

The JSNS$^2$ experiment at J-PARC has recently reported a high-statistics measurement of $\nu_\mu$ scattering on $^{12}$C using kaon decay at rest (KDAR) at $E_\nu\simeq235.5$~MeV\,\cite{ParticleDataGroup:2020ssz}. 
In contrast to earlier KDAR-based studies (e.g., MiniBooNE\,\cite{Mini}), the JSNS$^2$ result provides the differential distribution in the missing energy, $E_m$, thereby exposing how distinct nuclear mechanisms populate different regions of the spectrum. 
Because the beam energy is fixed, features of the measured $E_m$ distribution can be attributed largely to nuclear dynamics (binding energy, 2p--2h, and FSI), making this dataset one of the cleanest benchmarks for model testing in the sub-GeV domain of neutrino interactions\,\cite{Spitz:2014hwa}.

The original JSNS$^2$ presentation compared the measured $E_m$ distribution to several widely used event generators and RMF-based models, including \textsc{NuWro}\,\cite{Golan:2012wx,Golan:2012rfa,NuWro2019,NuWro2024}, \textsc{GiBUU}\,\cite{GiBUU2009,GiBUU2024}, and RMF+Achilles\,\cite{ACHILLES2022}. 
While all three reproduce the approximate position of the inelastic p-shell peak, they differ markedly in spectral width and in the strength redistributed into the high missing energy ($E_m$) tail where 2p--2h/FSI mechanisms become prominent \cite{Nico2021}. 
When confronted with the JSNS$^2$ spectrum, their reported goodness-of-fit values ($\chi^2\simeq35.5$ for \textsc{NuWro}, $176.8$ for \textsc{GiBUU}, and $58.1$ for RMF+Achilles) make clear that no single model simultaneously accounts for the peak width, the intermediate region, and the tail. 
These tensions point to simplified binding energy treatments, incomplete or implementation-dependent 2p--2h strength, and differing FSI transport pictures as the dominant sources of discrepancy. 
In short, the existing theory landscape offers valuable guidance but does not yet deliver a quantitatively faithful reproduction of the full spectral shape.

Motivated by this situation, we adopt a complementary, data-driven strategy: construct a minimal but physically interpretable empirical representation of the spectrum that (i) aligns with known nuclear structure expectations for the $p$- and s-shells and (ii) flexibly accommodates asymmetric broadening and continuum coupling induced by FSI and 2p--2h dynamics. 
To this end, each bound-shell component is modeled with an exponentially modified Gaussian (ex-Gaussian) line shape, and the continuum is encoded by a generalized power--exponential term. 
The ex-Gaussian form, long used to describe one-sided energy-loss or delayed-response processes, preserves an intrinsic Gaussian core while permitting controlled skewness set by an exponential scale $\tau$; it reduces continuously to a Gaussian when $\tau\!\to\!0$, avoiding unphysically inflated widths to mimic asymmetry. 
The continuum factor introduces an exponent parameter $\beta$ so that the slope can deviate from a pure exponential and match the measured falloff across the $E_m$ window of interest. 
This choice is not merely convenient: it mirrors the physics expectation that deep continuum strength arises from multinucleon knockout and intranuclear rescattering, which need not project onto a single exponential scale.

A second design principle is consistency with the experimental observable. 
All fits are performed in the visible-energy variable $E_{\rm vis}$ (the axis on which the JSNS$^2$ distribution is binned and uncertainties are quoted), with $E_m$ shown for presentation via a one-to-one mapping; this keeps the statistical treatment faithful to the measurement and avoids artifacts from bin redefinitions. 
Furthermore, we restrict the fitted range to the region where the detector response and selection do not induce threshold distortions (very low $E_{\rm vis}$ bins are excluded on these grounds), yielding a stable and physically interpretable result. 
Within this fitted energy range, the empirical form reproduces the measured spectrum with 
the chi-squared per number of degrees of freedom (ndf), \(\chi^2/\mathrm{ndf}\approx 1.3\), 
and the extracted parameters follow the anticipated pattern: a narrow, nearly symmetric p-shell peak, a broader s-shell with mild asymmetry consistent with continuum coupling, and a smooth bound-to-continuum transition encoded by the generalized tail. 
Although the model is empirical, each parameter has a transparent physical role (intrinsic widths, asymmetry scales, and a continuum falloff) and therefore provides a compact language for discussing how binding energy, 2p--2h, and FSI redistribute strength across the spectrum.

This empirical framework is intentionally modest in scope-it does not replace microscopic theory or transport calculations. 
Rather, it supplies a quantitatively faithful, model-independent representation of the JSNS$^2$ line shape that can be used as a common reference for phenomenological comparisons, cross-checks among generators, and future studies that seek to isolate the fingerprints of few-body correlations in inclusive observables. 
In particular, the minimal nature of the parameter set makes it straightforward to test whether proposed improvements in binding energy treatments, 2p--2h implementations, or FSI schemes actually shift the features (peak width, shoulder, and tail) in the directions required by data. 
We view this as complementary to generator development: empirical structure distilled from data can guide microscopic refinements without committing to a specific transport or correlation ansatz.

The remainder of the paper is organized as follows. 
Section~\ref{sec2} summarizes the dataset, the choice of observable, and the fitting methodology in $E_{\rm vis}$, including the mapping to $E_m$ and the fitted range. 
Section~\ref{sec3} presents generator-by-generator comparisons with \textsc{NuWro}, \textsc{GiBUU}, and RMF+Achilles and details the empirical fit performance across the spectrum, with emphasis on the intermediate and high-$E_m$ regions. 
Section~\ref{sec4} discusses the physical implications of the extracted parameters for QE and 2p--2h FSI dynamics and outlines how this representation can be used in future phenomenological studies. 
Section~\ref{sec5} provides conclusions.

\section{Observable, Data, and Analysis Setup}\label{sec2}
Our analysis uses the JSNS$^2$ measurement of the missing-energy spectrum in monoenergetic $\nu_\mu$ charged-current interactions on $^{12}$C at $E_\nu\simeq235.5$~MeV. Following the experimental convention, the missing energy is defined as the portion of the energy transfer to the nucleus subtracted by the outgoing proton(s), 
\begin{equation}
	E_m\;\equiv\omega-\sum_i T_{p,i},
\end{equation}
which, for the KDAR configuration, is related to the detector's visible energy by
\begin{equation}
	E_m\;=\;E_\nu - m_\mu + (m_n-m_p)\;-\;E_{\rm vis}\,,
\end{equation}
as given in the JSNS$^2$ publication (numerically $E_\nu=235.5~{\rm MeV}$, $m_\mu\simeq105.6584~{\rm MeV}$, $m_n-m_p\simeq1.3~{\rm MeV}$). This observable isolates nuclear effects (separation energy, Fermi motion, Pauli blocking, and FSI) that would otherwise be entangled with flux and kinematics, and it is naively expected to vanish in the absence of such effects. 

For fits and statistical comparisons we work in the visible-energy variable, $E_{\rm vis}$, which is the axis on which the JSNS$^2$ result is binned and for which uncertainties are reported, and display the corresponding $E_m$ through the one-to-one mapping. In our implementation we adopt the precise conversion provided in the JSNS$^2$ data release,
\begin{equation}
	E_m\;=\;129.8736~{\rm MeV}\;-\;E_{\rm vis}\,,
\end{equation}
where the constant follows from $E_\nu=235.532~{\rm MeV}$ and $m_\mu=105.6584~{\rm MeV}$ used in the release. All shape-only comparisons are invariant under this linear change of variable.

The experimental inputs (binned spectrum in $E_{\rm vis}$ and the full covariance) are taken directly from the JSNS$^2$ PRL supplementary materials/data release and are used in all data-model tests and in our empirical fit so that bin-to-bin correlations are propagated consistently\,\cite{JSNS2}. We restrict the fit window to the region where the detector response is well-behaved (approximately $5~{\rm MeV} < E_m < 85~{\rm MeV}$), following the published selection\,\cite{JSNS2}. Because only the spectral shape is accessible, every spectrum (data or model) is normalized to unit area over the common range; absolute cross sections are not used. 

For the three reference predictions highlighted by JSNS$^2$-\textsc{NuWro} (v21, $^{12}$C spectral function) \cite{NuWro2024,NuWro2019} \textsc{GiBUU} (2021p1) \cite{GiBUU2009,GiBUU2024}, and RMF+Achilles \cite{ACHILLES2022}, we obtain bin-centered values by digitizing the published histograms so that they share the experimental binning. When overlaid on the data, comparisons use the experimental covariance; generator-side statistical fluctuations are not propagated in the overlay. The goodness-of-fit values we quote for these generators are the ones reported by JSNS$^2$ (in the higher-statistics window used there), namely $\chi^2\simeq35.5$ (\textsc{NuWro}), $176.8$ (\textsc{GiBUU}), and $58.1$ (RMF+Achilles), and readers are referred to the JSNS$^2$ analysis for generator configurations and unfolding details.

In parallel with these overlays, we fit a compact empirical representation of the spectrum in which the bound-shell contributions ($p$ and $s$) are modeled by exponentially modified Gaussian (ex-Gaussian) line shapes and the continuum is described by a generalized power-exponential term. This construction preserves intrinsic widths while allowing controlled asymmetry (to encode energy-loss/FSI effects) and a flexible falloff of the continuum (to encode multi-nucleon and rescattering strength), enabling a data-driven decomposition of the peak region, intermediate shoulder, and high-$E_m$ tail within a single, physically interpretable framework. 

\section{Results}\label{sec3}

\subsection{Overall reproduction of the JSNS$^2$ spectrum}\label{subsec:repro}
Using the official JSNS$^2$ binned $E_{\rm vis}$ spectrum and its covariance (Sec.~\ref{sec2}), and the linear $E_{\rm vis}\!\leftrightarrow\!E_m$ mapping defined therein, we reconstructed the distribution within the common analysis window $20~\mathrm{MeV}<E_{\rm vis}<150~\mathrm{MeV}$ (or $-20$ MeV $< E_m <$ 110 MeV). All spectra are area-normalized over this window, and bin-to-bin correlations are propagated with the provided covariance. The reconstructed result agrees with the published spectrum within the quoted uncertainties across the fitted range; the residuals are consistent with zero and do not exhibit a visible systematic trend. This reference spectrum is used as the baseline for generator overlays (Sec.~\ref{subsec:gens}), for the focused discussion of the high-$E_m$ region (Sec.~\ref{subsec:tail}), and as the target of the empirical fit (Sec.~\ref{subsec:exg}).

\subsection{Generator-by-generator comparisons}\label{subsec:gens}
To assess the behavior of different theoretical descriptions, we compare each generator prediction with the JSNS$^2$ spectrum on the experimental binning and within the common analysis window. For every case we show (i) the area--normalized differential distribution (top), (ii) residuals defined as $\Delta(\mathrm{Data}-\mathrm{Model})$ on the experimental bins (middle), and (iii) the data--to--model ratio (bottom). In the middle and bottom panels, gray bands depict statistical uncertainties and red bands indicate statistical$\oplus$systematic uncertainties propagated from the experimental covariance. This panel layout makes the regions of agreement and discrepancy immediately visible without relying on model--dependent rebinning. 
Using the published covariance and the 16-bin window ($E_m$ = 5--85 MeV), we independently recomputed $\chi^2$ for the three references and confirmed agreement with the JSNS${^2}$ results.

\begin{figure*}
	\centering
	\includegraphics[width=0.78\textwidth]{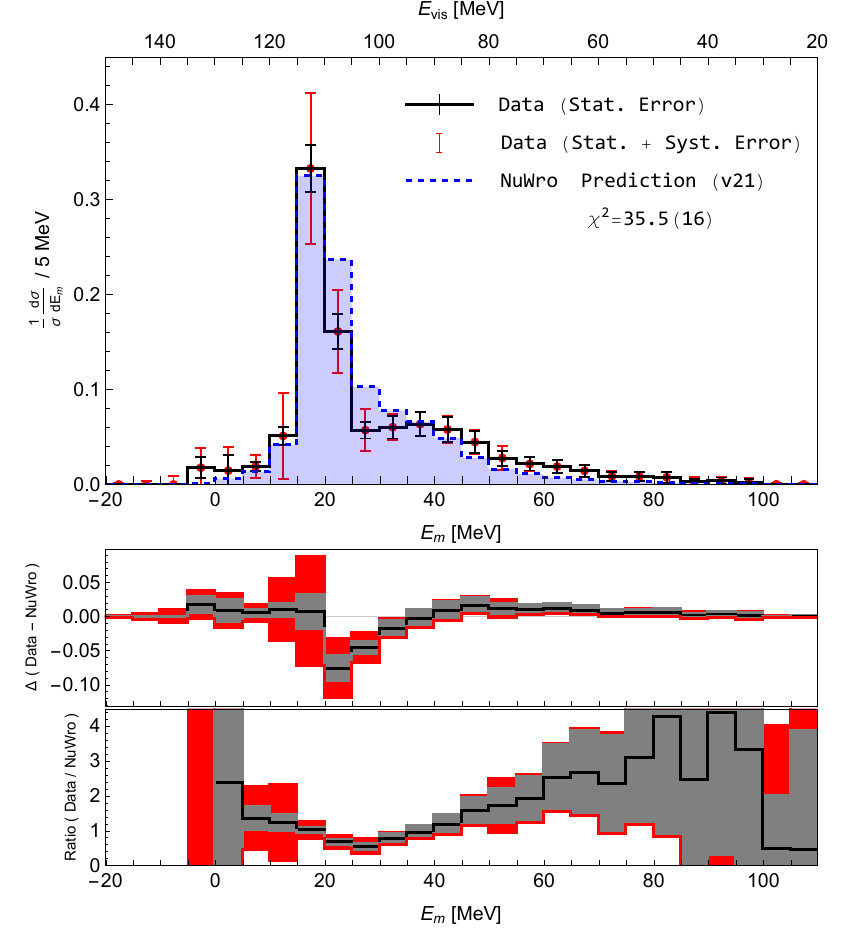}
     \caption{Comparison of the JSNS$^2$ missing--energy spectrum with the \textsc{NuWro} (v21) prediction.
	Top: area--normalized differential distribution per 5~MeV bins; black bars show statistical errors and red bars indicate statistical$\oplus$systematic uncertainties. The dashed blue histogram is the \textsc{NuWro} result. The upper axis labels $E_{\rm vis}$, while the lower axis labels $E_m$.
	Middle: bin--by--bin residuals $\Delta(\text{Data}-\text{NuWro})$ on the experimental binning.
	Bottom: data--to--model ratio.
	\textsc{NuWro} reproduces the peak location but does not capture the observed peak width and underestimates the strength in the intermediate-$E_m$ region and in the high-$E_m$ tail; 
	the corresponding global comparison yields $\chi^2=35.5$ for 16 degrees of freedom.}
	\label{fig:nuwro}
\end{figure*}	
	\begin{figure*}
	\centering
	\includegraphics[width=0.78\textwidth]{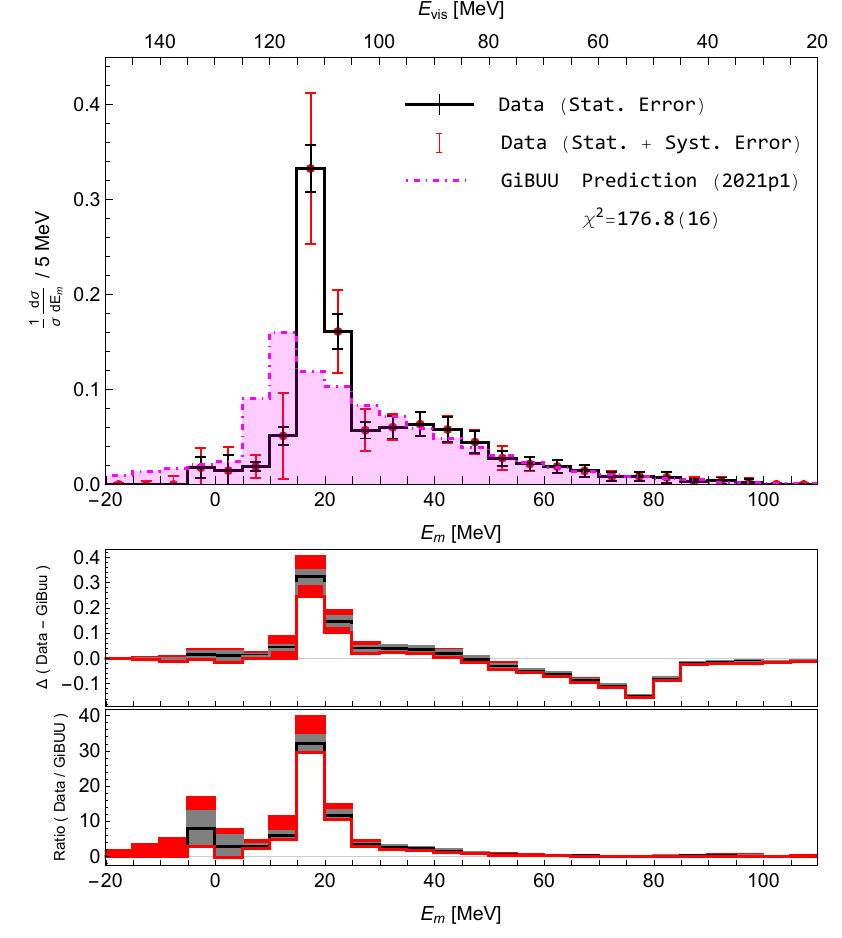}
\caption{As in Fig.~1 (same panel layout and area normalization), but for \textsc{GiBUU} (2021p1). 
	The magenta histogram shows the \textsc{GiBUU} prediction. 
	Relative to the data, \textsc{GiBUU} overshoots strength in the low-- and intermediate--$E_m$ regions and shifts yield toward higher $E_m$, resulting in a large global comparison value $\chi^2=176.8$ for 16 degrees of freedom.}
	\label{fig:gibuu}
\end{figure*}
\paragraph*{\textsc{NuWro}.}
Figure~\ref{fig:nuwro} contrasts the JSNS$^2$ spectrum with the \textsc{NuWro} prediction (v21). The model reproduces the peak location but does not capture the observed peak width, and it underestimates strength in both the intermediate--$E_m$ region and the high--$E_m$ tail. The residuals exhibit coherent negative excursions in the $E_m$ $\sim$ 30--50~MeV interval and remain negative into the tail, consistent with the data--to--model ratio falling below unity in these regions. The global comparison reported by JSNS$^2$ yields $\chi^2=35.5$ for 16 degrees of freedom, indicating a shape tension concentrated outside the peak region.
NuWro is a lightweight and flexible neutrino event generator developed by the Warsaw group \cite{NuWro2019,NuWro2024}. 
It encompasses several nuclear models, such as the relativistic Fermi gas (RFG), local Fermi gas (LFG), and spectral functions, allowing users to test different nuclear descriptions easily. Quasi-elastic scattering, resonance production, deep inelastic scattering, and 2p--2h contributions are also included. Final-state interactions (FSI) are treated with a simplified intranuclear cascade. Its main strength lies in being fast, easy to use, and adaptable for theoretical studies and comparisons with data, although its FSI modelling is treated in a simple way compared to more advanced approaches.
In brief, NuWro is best for flexibility and quick theoretical and experimental comparisons\,\cite{Golan:2012wx,Golan:2012rfa}. 
But, as shown Fig.~\ref{fig:nuwro}, the high energy tail part is a bit underestimated due to the simple FSI modelling despite of the recent development of the cascade model  by the Monte Carlo method \cite{NuWro2019}. Another mismatch is found in the right shoulder of the peak, which corresponds to the overlap energy region of the low-energy inelastic scattering and the QE neutrino scattering region.

\paragraph*{\textsc{GiBUU}.}
Figure~\ref{fig:gibuu} compares the spectrum with \textsc{GiBUU} (2021p1). The calculation redistributes strength toward larger $E_m$, leading to pronounced positive residuals at low-- and intermediate--$E_m$ and a data--to--model ratio below unity over a broad range; together these features quantify the overshoot in those regions. The reported global comparison is $\chi^2=176.8$ for 16 degrees of freedom, consistent with sizable shape tension driven by transport--induced migration of yield from the peak into higher $E_m$. 
GiBUU, developed at Giessen University, is based on transport theory and solves the Boltzmann-Uehling-Uhlenbeck (BUU) equation for particle propagation in the nuclear medium \cite{GiBUU2009,GiBUU2024}. It covers a wide range of processes, including quasi-elastic, resonance, DIS, coherent, and multi-nucleon interactions. GiBUU provides the most detailed description of FSI, since produced particles are dynamically transported through the nucleus with absorption and rescattering explicitly modelled. This makes it the most reliable tool for precision studies of nuclear effects, but it is computationally demanding and less convenient for large-scale event generation compared to NuWro or GENIE.
In brief, GiBUU excels in detailed FSI and transport dynamics, making it the most precise but also the heaviest computationally. As shown Fig.~\ref{fig:gibuu}, the high energy tail is well reproduced, but the peak strength is underestimated, and the peal position is shifted by about 5 MeV. It means that the GiBUU is focused on the QE scattering region by the transport theory\,\cite{Lalakulich:2011eh, Buss:2011mx}.

\begin{figure*}[t]
	\centering
	\includegraphics[width=0.8\textwidth]{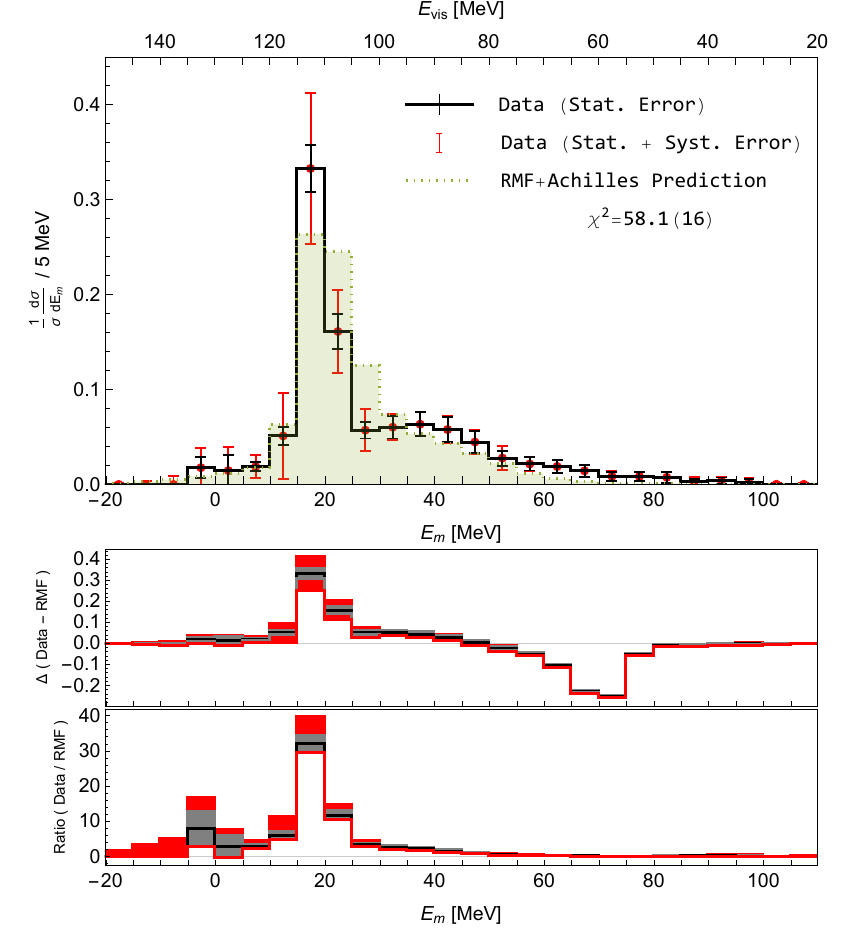}
\caption{As in Fig.~1 (same panel layout and area normalization), but for RMF+Achilles (green histogram).
	The model reproduces the peak location and captures part of the high--$E_m$ tail, while discrepancies remain in the detailed shape around $E_m\!\sim\!20$--30~MeV and in the peak width; the global comparison yields $\chi^2=58.1$ for 16 degrees of freedom.}
	\label{fig:rmf}
\end{figure*}
\paragraph*{RMF+Achilles.}
Figure~\ref{fig:rmf} shows the comparison with RMF+Achilles. The model aligns with the peak location and captures part of the high--$E_m$ tail, but residual differences appear in the $E_m\!\sim\!20$--30~MeV region and in the peak strength width, as seen from structured residuals and a ratio that departs from unity in those bins. The corresponding global value is $\chi^2=58.1$ for 16 degrees of freedom, indicating improved but still incomplete agreement across the full shape. 
RMF+ACHILLES (A CHIcagoL and Lepton Event Simulator) is a more recent event generator motivated by the relativistic mean field (RMF) approach and superscaling analyses \cite{ACHILLES2022}. Its strength becomes significant in the quasi-elastic region, where nucleon dynamics are described self-consistently with RMF wave functions, including off-shell effects. It also incorporates superscaling functions from electron scattering (SuSA/SuSAv2) \cite{SuSA2020}, ensuring consistency with experimental data. Final-state interactions are described at the mean-field level, rather than through full transport. As a result, RMF+Achilles provides a theoretically robust description of QE scattering and is particularly suitable to intermediate-energy neutrinos, such as KDAR neutrinos around 236 MeV. However, it lacks the broad coverage of high-energy processes and detailed transport modelling found in GiBUU.
In brief, RMF+ACHILLES provides the strongest theoretical foundation for QE scattering, especially in the sub-GeV regime, though with limited scope compared to the other two generators. Fig.~\ref{fig:rmf} shows a good result in the high energy tail region than those by NuWRO prediction, but still lacks of accuracy compared to those by GiBUU. It may come from that the QE part, {\it i.e.}, the high energy tail is only scaled to the QE electron scattering data \cite{Kim2022}. Moreover, the peak strength is still underestimated because of the simple RMF approach\,\cite{Gonzalez-Jimenez:2019qhq, Benhar:1994hw, Gonzalez-Jimenez:2021ohu, ACHILLES2022, Isaacson:2020wlx}.

\subsection{High--missing-energy tail}\label{subsec:tail}
We place particular emphasis on the high--$E_m$ region ($\gtrsim$\,40~MeV), where the generator predictions diverge most visibly in the residual and ratio panels. This part of the spectrum is driven by multinucleon (2p--2h) dynamics and by intranuclear rescattering, so modest differences in the modelling of these ingredients translate into large shape changes. In our overlays, \textsc{NuWro} systematically underpredicts the strength beyond the peak shoulder, \textsc{GiBUU} redistributes yield toward higher $E_m$ and overshoots over a broad interval, and RMF+Achilles captures part of the tail but still departs from the data in detail. Because all spectra are area--normalized and the experimental covariance is propagated, these trends reflect genuine shape differences rather than overall normalization effects. The high--$E_m$ tail therefore provides a discriminating test of nuclear dynamics beyond simple quasielastic scattering and motivates a flexible continuum component in the empirical description introduced below (Sec.~\ref{subsec:exg}).

\subsection{Empirical Ex-Gaussian Decomposition of the Spectrum}\label{subsec:exg}
All analyses and parameter optimizations were performed directly in the visible-energy variable,
$E_{\mathrm{vis}}$, which defines the experimental binning and statistical uncertainties of the JSNS$^{2}$ KDAR data.
This ensures that the fitting procedure is fully consistent with the detector response and measured observables.
For comparison with previous missing-energy representations, the relation
$E_{\mathrm{miss}} = 129.87 - E_{\mathrm{vis}}$
can be used, but all model parameters quoted here correspond to the fitting in $E_{\mathrm{vis}}$ space.

While the first analysis in Ref.~\cite{JSNS2} employed a phenomenological
two-Gaussian plus exponential-tail form to describe the spectrum,
such a symmetric parametrization cannot fully reproduce the observed shape
once the detector resolution and nuclear asymmetries are taken into account.
In particular, the s-shell region of the JSNS$^2$ KDAR spectrum
appears broader and more asymmetric than expected from a simple Gaussian form.

To address this limitation, we adopt a physically motivated 
\emph{two ex-Gaussian plus tail} model.
Each bound-state contribution (p- and s-shell) is represented by an
ex-Gaussian line shape the convolution of a Gaussian core and an exponential tail that captures asymmetric broadening induced by final-state interactions (FSI),
multi-nucleon (2p--2h) coupling, and coupling to the continuum.
A detailed derivation of this functional form and its normalization
is provided in Appendix~\ref{secA1}.
This form has long been employed in detector and nuclear physics
to describe one-sided energy-loss processes and delayed responses,
and provides a natural extension of the symmetric Gaussian model.
The ex-Gaussian preserves the intrinsic shell width while allowing for
finite skewness governed by an asymmetry scale~$\tau$,
where positive~$\tau$ produces a low-$E_{\mathrm{vis}}$ tail,
corresponding to an energy-loss process.
In the limit $\tau\!\to\!0$, the function smoothly reduces to a
pure Gaussian, eliminating the need for unphysically broad widths
to mimic asymmetric features.

\begin{figure*}
	\centering
	\includegraphics[width=1.0\textwidth]{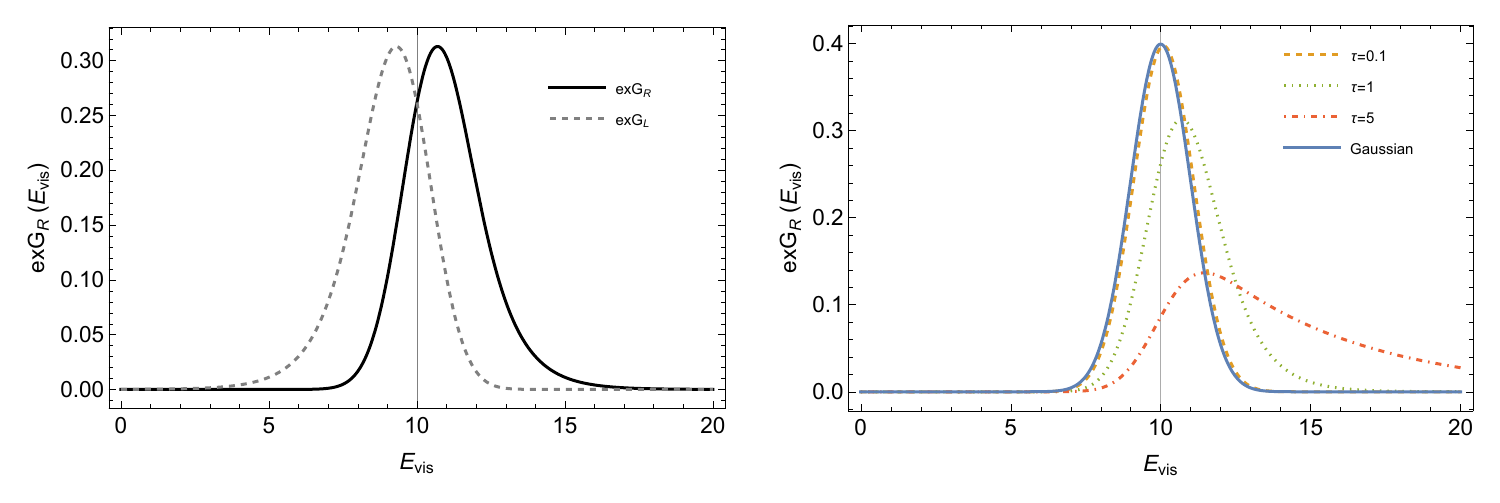}
\caption{
	Illustration of the ex-Gaussian line shape used to describe asymmetric 
	broadening in the missing-energy spectrum.
	(\textbf{Left}) Comparison between the right- and left-skewed forms,
	$\mathrm{exG_R}$ (solid) and $\mathrm{exG_L}$ (dashed),
	which are mirror reflections about the mean value 
	($\mathrm{exG_L}(E_m)=\mathrm{exG_R}(2\mu-E_m)$). 
	(\textbf{Right}) Dependence of $\mathrm{exG_R}$ on the asymmetry scale $\tau$,
	together with the Gaussian limit (solid blue curve).
	Smaller $\tau$ values lead to a nearly symmetric Gaussian-like shape,
	whereas larger $\tau$ produce an extended high-$E_m$ tail.
	This behavior enables the ex-Gaussian to model realistic one-sided
	broadening due to final-state interactions and continuum coupling.
}
\label{FigExGaussian}
\end{figure*}

To illustrate these properties, Fig.~\ref{FigExGaussian} shows 
several representative examples of the ex-Gaussian line shape.
The function is defined as the convolution of a Gaussian core and an exponential tail,
which produces an asymmetric broadening that cannot be reproduced by a simple Gaussian.
Two functional forms are used depending on the direction of the tail:
the right-skewed form, ${\rm exG_R}$, corresponds to a positive asymmetry scale 
($\tau>0$) and generates a high-$E_{\mathrm{vis}}$ tail,
whereas the left-skewed form, ${\rm exG_L}$, with $\tau<0$, 
represents the mirror-symmetric shape extending toward low~$E_{\mathrm{vis}}$.
Their explicit expressions are
\begin{widetext}
\begin{eqnarray}
	{\rm exG_R}(E_{\mathrm{vis}};\mu,\sigma,\tau)
	&=& \frac{1}{2\tau}\,
	\exp\!\left[\frac{\sigma^2}{2\tau^2}-\frac{E_{\mathrm{vis}}-\mu}{\tau}\right]\,
	\mathrm{erfc}\!\left(
	\frac{\sigma}{\sqrt{2}\tau}-\frac{E_{\mathrm{vis}}-\mu}{\sqrt{2}\sigma}
	\right), \\[4pt]
	{\rm exG_L}(E_{\mathrm{vis}};\mu,\sigma,\tau)
	&=& \frac{1}{2\tau}\,
	\exp\!\left[\frac{\sigma^2}{2\tau^2}+\frac{E_{\mathrm{vis}}-\mu}{\tau}\right]\,
	\mathrm{erfc}\!\left(
	\frac{\sigma}{\sqrt{2}\tau}+\frac{E_{\mathrm{vis}}-\mu}{\sqrt{2}\sigma}
	\right),
\end{eqnarray}
\end{widetext}
where $\mathrm{erfc}(x)=1-\mathrm{erf}(x)$ is the complementary error function. 
As shown in the left panel of Fig.~\ref{FigExGaussian}, 
the two forms are related by reflection about the mean value,
${\rm exG_L}(E_{\mathrm{vis}})={\rm exG_R}(2\mu-E_{\mathrm{vis}})$.
The right panel illustrates how the ex-Gaussian evolves with varying $\tau$:
as $\tau$ decreases, the distribution gradually converges to the 
symmetric Gaussian limit (solid blue curve),
while larger $\tau$ values enhance the one-sided exponential tail 
toward lower~$E_{\mathrm{vis}}$,
representing stronger energy-loss or delayed-response components.
This behavior demonstrates how the ex-Gaussian provides a simple yet 
physically meaningful representation of the asymmetric line shapes 
caused by FSI, 2p--2h processes, and continuum coupling, 
while retaining the intrinsic width of each nuclear shell.

\paragraph{Fit model.}
The total fit model used in this analysis is expressed in the visible-energy domain as
\begin{widetext}
\begin{equation}
	F(E_{\mathrm{vis}}) \;=\;
	A_P\,{\rm exG_L}(E_{\mathrm{vis}};\mu_P,\sigma_P,\tau_P)
	+ A_S\,{\rm exG_L}(E_{\mathrm{vis}};\mu_S,\sigma_S,\tau_S)
	+ A_T\,T(E_{\mathrm{vis}};E_0,\lambda,\beta),
\end{equation}
\end{widetext}
where \(A_P\), \(A_S\), and \(A_T\) are fit parameters controlling the amplitudes of the p-shell, s-shell, and continuum components, respectively. The bound-state shapes use \(\mathrm{exG}_L\big(E_{\mathrm{vis}};\mu,\sigma,\tau\big)\), where \(\mathrm{exG}_L\) denotes the normalized left-skewed ex-Gaussian function, corresponding to an energy-loss tail toward lower \(E_{\mathrm{vis}}\). The continuum term \(T(E_{\mathrm{vis}};E_0,\lambda,\beta)\) is governed by the onset \(E_0\), the upper bound \(E_t\), the falloff scale \(\lambda\), and the exponent \(\beta\).

\begin{figure*}
	\centering
	\includegraphics[width=1.00\textwidth]{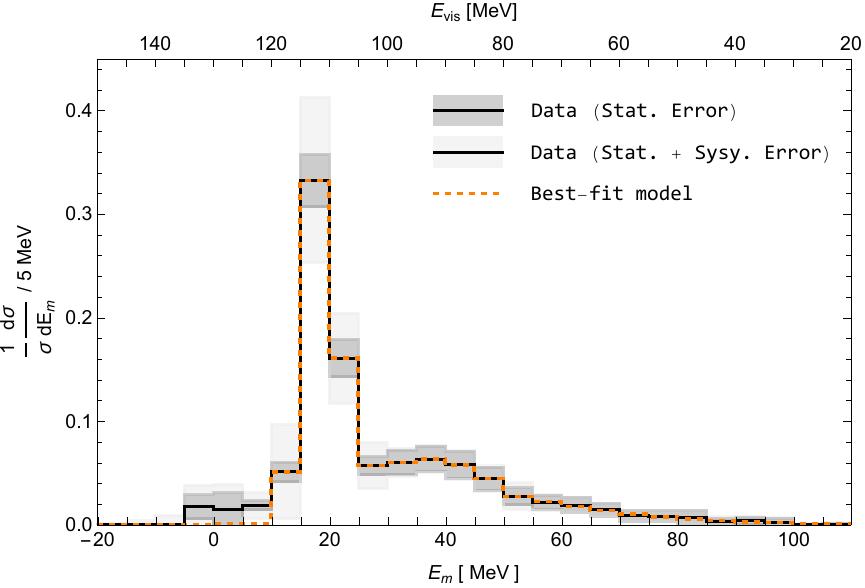}
	\caption{Comparison between the JSNS$^2$ missing-energy spectrum and the best-fit model. 
		The black histogram shows the experimental data, with dark and light gray bands 
		representing the statistical and total (statistical + systematic) uncertainties, respectively. 
		The orange dashed line denotes the best-fit model averaged over each experimental bin width. 
		Within the fitted region of $10~\mathrm{MeV} < E_m < 120~\mathrm{MeV}$, 
		the model closely follows the central values of the measured spectrum, 
		reproducing both the p-shell and s-shell features as well as the overall 
		high-$E_m$ continuum within uncertainties.
	}
	\label{fig:exG_fit}
\end{figure*}

Complementing the two ex-Gaussian bound-state terms, 
the third term T provides a generalized power-exponential continuum that captures strength from multi-nucleon (2p-2h) dynamics, short-range correlations (SRC), and FSI-induced migration:
\begin{widetext}
\begin{equation}
T(E_{\mathrm{vis}};\lambda,\beta)
= \exp\!\left[-\left(\frac{E_t-E_{\mathrm{vis}}}{\lambda}\right)^{\!\beta}\right]
\Theta(E_{\mathrm{t}}-E_{\mathrm{vis}})\,\Theta(E_{\mathrm{vis}}-E_0),
\end{equation}
\end{widetext}
where $E_0$ and $E_{\mathrm{t}}$ define the onset and upper boundary of the fitted
continuum region, and $\Theta(x)$ is the Heaviside step function
(equal to $1$ for $x>0$ and $0$ for $x<0$).
The additional exponent parameter $\beta$ allows the continuum slope to deviate
from a pure exponential form ($\beta=1$),
providing greater flexibility to reproduce the experimental spectrum 
in the visible-energy range ($E_0 \!\lesssim\! E_{\mathrm{vis}} \!\lesssim\! E_{\mathrm{t}}$).
The optimized parameters are found to be
$E_0 \simeq 20~\mathrm{MeV}$,
$E_{\mathrm{t}} \simeq 78~\mathrm{MeV}$,
$\lambda \simeq 21~\mathrm{MeV}$,
and $\beta \simeq 1.24$.
This generalized tail representation describes the smooth transition from the
continuum to the bound-shell region with improved accuracy.
The resulting fit reproduces the measured $E_{\mathrm{vis}}$ spectrum with excellent
agreement, as shown in Fig.~\ref{fig:exG_fit}.

The fit range was restricted to
$20~\mathrm{MeV}<E_{\mathrm{vis}}<122~\mathrm{MeV}$,
covering the full p- and s-shell regions.
The very low-energy domain ($E_{\mathrm{vis}}\le 20~\mathrm{MeV}$)
was excluded because it is dominated by low-lying inelastic events and threshold
effects that distort the reconstructed visible energy.
Including this region leads to systematic bias and an artificial increase in
$\chi^2$.
By excluding it, the fit remains physically meaningful,
yielding a substantially improved $\chi^2/{\mathrm{ndf}}$ and a
consistent description of the nuclear shell structure.

\paragraph{Fit result.}
The best-fit parameters obtained from the JSNS$^2$ KDAR spectrum are
\begin{widetext}
\[
\begin{aligned}
	A_S&=2.01, &\quad 
	\mu_S&=93.55~\mathrm{MeV}\;(\,\mu_S^{E_m}=36.32~\mathrm{MeV}\,), &\quad
	\sigma_S&=12.58~\mathrm{MeV}, &\quad 
	\tau_S&=0.49~\mathrm{MeV},\\[3pt]
	A_P&=2.46, &\quad 
	\mu_P&=113.08~\mathrm{MeV}\;(\,\mu_P^{E_m}=16.79~\mathrm{MeV}\,), &\quad
	\sigma_P&=2.65~\mathrm{MeV}, &\quad 
	\tau_P&=1.67~\mathrm{MeV},
\end{aligned}
\]
\end{widetext}
where \(\mu^{Em}_S\) and \(\mu^{Em}_P\) denote, respectively, the s-shell and p-shell peak positions expressed on the missing-energy scale.
The resulting $\chi^2/{\mathrm{ndf}}\!\simeq\!1.3$ indicates an excellent
agreement between the model and data across the entire fit range
($5~\mathrm{MeV}<E_{\mathrm{miss}}<85~\mathrm{MeV}$ or $44.8 < E_{\mathrm{vis}} < 124.9 ~\mathrm{MeV}$).
\footnote{%
	All $\chi^2$ values were computed using the full experimental covariance matrix 
	(statistical and systematic) as 
	$\chi^2=(d-p)^{\mathrm T}V^{-1}(d-p)$, 
	restricted to the 16 bins corresponding to $E_m=5$--$85~\mathrm{MeV}$. 
	In the original JSNS$^2$ analysis~\cite{JSNS2}, 
	no model parameters were fitted ($N_{\mathrm{param}}=0$), 
	so $N_{\mathrm{dof}}=16$ was used, giving 
	$\chi^2/\mathrm{dof}=2.22$, $11.0$, and $3.63$ for 
	NuWro, GiBUU, and RMF+Achilles, respectively. 
	Applying the same convention to our empirical ex-Gaussian model 
	yields $\chi^2/16\simeq0.5$. 
	When the ten fit parameters 
	($A_{P,S}$, $\mu_{P,S}$, $\sigma_{P,S}$, $\tau_{P,S}$, $\lambda$, $\beta$) 
	are counted explicitly, the effective degrees of freedom become 
	$N_{\mathrm{dof}}=16-10=6$, corresponding to 
	$\chi^2/\mathrm{dof}\simeq1.3$, confirming an excellent agreement 
	between the model and data.%
}
The fitted amplitudes and widths are well constrained, with negligible
parameter degeneracy, demonstrating that the adopted two--ex-Gaussian
model provides a unique and physically consistent decomposition of the
measured spectrum.

\paragraph{Fit summary.}
The spectrum is represented by two bound-state components
(modeled with skewed Gaussian functions) plus a smooth continuum.
The p-shell peak is characterized by
$\mu_P\simeq113~\mathrm{MeV}$,
$\sigma_P\simeq2.7~\mathrm{MeV}$,
and $\tau_P\simeq1.7~\mathrm{MeV}$,
while the s-shell peak has
$\mu_S\simeq94~\mathrm{MeV}$,
$\sigma_S\simeq12.6~\mathrm{MeV}$,
and $\tau_S\simeq0.5~\mathrm{MeV}$.
The centroid separation of $\Delta\mu\approx19~\mathrm{MeV}$
is consistent with the empirical $1p$--$1s$ gap in carbon.
The quoted s-shell width represents an \emph{effective} width that
includes unresolved continuum strength and detector smearing\,\cite{NuWro2024, Ankowski:2024ntv}.

\section{Discussion and Physical Implications}\label{sec4}
The present analysis demonstrates that a symmetric double-Gaussian description is insufficient
to reproduce the measured JSNS$^2$ KDAR missing-energy spectrum.
Earlier fits required $\sigma_S\!\gtrsim\!14$~MeV to absorb the asymmetric continuum strength,
suggesting that part of the non-Gaussian behavior was being folded into the s-shell component.
Introducing an ex-Gaussian line shape for both shells resolves this issue.
The additional asymmetry parameters $\tau_S$ and $\tau_P$ capture the local skewness caused by
final-state interactions (FSI) and coupling to the continuum,
allowing the intrinsic widths to stabilize at physically reasonable values
($\sigma_S$ ${\simeq}$ 12.6~MeV, $\sigma_P$ ${\simeq}$ 2.7~MeV)
without compromising fit quality.
At the current statistical precision, further subdivision of the s-shell
(e.g., separating explicit 2p--2h or FSI subcomponents)
does not yield a statistically significant improvement,
indicating that the adopted two--ex-Gaussian plus tail form
is the most economical and self-consistent description achievable with existing data.

The decomposition reveals a clear dominance by the two bound-state peaks correspond to the p- and s-shell knockouts,
with a smooth continuum term bridges the transition to higher missing energy.
The p-shell remains narrow and nearly symmetric,
reflecting surface knockout with minimal rescattering.
In contrast, the s-shell exhibits a broader, mildly asymmetric shape,
consistent with core nucleon removal strongly coupled to continuum states.
The $\sim$20~MeV separation between $\mu_P$ and $\mu_S$ in Eqs.(4) and (5)
is compatible with the empirical $1p$--$1s$ single-particle energy gap in carbon,
supporting the physical shell assignment of the two peaks.
The large effective width of the s-shell should therefore be interpreted
as the convolution of intrinsic bound strength with unresolved
continuum and detector-resolution effects, rather than the intrinsic bound-state width itself.

In the intermediate region ($E_m$ $\approx$ 30--50~MeV),
the extracted asymmetry scale $\tau_S$ and continuum slope parameters
imply that 2p--2h and SRC processes are dynamically coupled
to the s-shell domain, forming the asymmetric shoulder observed in the data.
This interpretation is consistent with $(e,e'p)$ and $(p,2p)$
spectral-function measurements on $^{12}$C,
where $s$-hole states display strong mixing with the continuum
while $p$-hole states remain relatively sharp and surface localized\,\cite{Makins1994}.

At higher $E_m$ ($\gtrsim$70~MeV),
the fitted power-exponential tail with $\lambda{\simeq}21$~MeV and $\beta{\simeq}1.24$
quantifies the smooth transition between the bound and deep continuum regimes.
This ``bridge'' component likely reflects the onset of
multinucleon knockout and inelastic rescattering.
Its magnitude and slope are sensitive to the strength of FSI,
underscoring the need for theoretical models that treat
the bound-to-continuum coupling consistently.

Overall, the ex-Gaussian decomposition provides a concise and physically transparent
representation of the KDAR missing-energy spectrum.
It captures, within a minimal set of parameters,
the essential nuclear response features:
(i) a narrow, symmetric p-shell peak,
(ii) a broader, asymmetric s-shell shaped by continuum coupling, and
(iii) a smooth high-$E_m$ continuum governed by FSI and multinucleon effects.
This result offers quantitative, data-driven evidence that
the s-shell response in carbon exhibits asymmetric broadening due to
continuum coupling, while the p-shell remains surface-dominated and
nearly Gaussian.
As such, the JSNS$^2$ measurement provides a stringent benchmark for
nuclear models aiming to unify quasielastic and 2p--2h dynamics
in the few-hundred-MeV neutrino energy regime.

\section{Conclusions}\label{sec5}
We have revisited the JSNS$^2$ measurement of muon--neutrino scattering on $^{12}$C
at 235.5~MeV, focusing on the missing-energy ($E_m$) spectrum as a sensitive probe of
nuclear response dynamics. Using the published distribution, we performed a detailed
shape-only analysis combining generator-by-generator comparisons with an empirical
decomposition based on physically motivated ex-Gaussian line shapes.

In the first part of this study, the JSNS$^2$ spectrum was compared individually
with several theoretical models and event generators, including NuWro, GiBUU,
and RMF+Achilles. This generator-by-generator approach revealed how specific nuclear
ingredients manifest in different regions of the spectrum.  
The peak region ($E_m \sim 20-30$~MeV) is primarily governed by binding energy
treatments and Fermi motion, while the intermediate region ($E_m \sim 30-50$~MeV) is shaped by
two-particle-two-hole (2p-2h) excitations and short-range correlations.
The high-$E_m$ tail ($\gtrsim$ 70~MeV) is dominated by FSI
and inelastic processes, providing the most powerful discriminator among competing
descriptions. These comparisons confirm that the JSNS$^2$ spectrum, thanks to its
monoenergetic KDAR neutrino source, serves as a precise benchmark for testing and
refining nuclear models.

Building on these insights, we introduced a data-driven \emph{two ex-Gaussian plus tail}
model to describe the experimental spectrum.  
Each bound-state contribution (p- and s-shell) was represented by an ex-Gaussian
line shape-the convolution of a Gaussian core and an exponential tail-while the
continuum was modelled by a generalized power-exponential function.
The resulting fit ($\chi^2/{\mathrm{ndf}}\!\simeq\!1.3$) achieved excellent agreement with
the data across the full range of $E_m\sim$ 10-110~MeV.  
The fitted parameters,
$\mu_P$ ${\simeq}$ 113~MeV, $\sigma_P$ ${\simeq}$ 2.7~MeV, $\tau_P$ ${\simeq}$ 1.7~MeV, and
$\mu_S$ ${\simeq}$ 94~MeV, $\sigma_S$ ${\simeq}$ 12.6~MeV, $\tau_S$ ${\simeq}$ 0.5~MeV,
together with $\lambda$ ${\simeq}$ 21~MeV and $\beta$ ${\simeq}$ 1.24 for the
continuum, reproduce the entire spectral shape with physically meaningful values.

This decomposition reveals a clear quasielastic dominance with two bound-state peaks
corresponding to p- and s-shell knockouts, and a smooth continuum that bridges
the transition to higher $E_m$.  
The narrow p-shell peak corresponds to surface proton knockout with weak FSI,
while the broader, mildly asymmetric s-shell reflects deeper nucleon removal strongly
coupled to the continuum.  
The $\sim$20~MeV separation between $\mu_P$ and $\mu_S$ matches the empirical
$1p$-$1s$ single-particle gap in carbon, providing confidence in the shell assignment.
The effective s-shell width ($\sigma_S^{\rm eff}\!\approx\!12.6$~MeV) should be viewed
as a convolution of intrinsic bound strength with unresolved 2p--2h and FSI effects,
rather than as an intrinsic width alone.

Taken together, these findings establish the JSNS$^2$ KDAR measurement as a
stringent testbed for neutrino-nucleus interaction models.  
The ex-Gaussian analysis provides the first quantitative, data-driven evidence that
the s-shell response in carbon exhibits asymmetric broadening from continuum
coupling, while the p-shell remains nearly Gaussian and surface-dominated.
Future improvements in binding energy treatments, multinucleon dynamics,
and FSI modelling-guided by this observable-will be essential for achieving a
unified and predictive description of quasielastic and 2p--2h processes in the
few-hundred-MeV regime\,\cite{JSNS2:2020hmg}.



\begin{acknowledgments}
The author thanks colleagues in the JSNS2 Collaboration for making the experimental results publicly available and for helpful discussions. 
This work was supported by grants from the National Research Foundation (NRF) of the Korean government
(RS-2021-NR060129, RS-2025-16071941, RS-2022-NR069287, RS-2022-NR070836, RS-2025-25400847, RS-2025-16071941, and RS-2025-24533596).
\end{acknowledgments}







\section*{DATA AVAILABILITY}
All data analyzed in this study are available from the published JSNS$^2$ results. 
Digitized spectra derived from these results and supplementary data used in the fitting 
can be obtained from the corresponding author upon reasonable request. \\



\appendix

\section{Derivation of the Ex-Gaussian Function}\label{secA1}
\appendix
The ex-Gaussian (exponentially modified Gaussian) distribution is obtained by
the convolution of a Gaussian function and an exponential decay.
Let $G \sim \mathcal{N}(\mu,\sigma^{2})$ and $E \sim \mathrm{Exp}(\tau)$ be
independent random variables, where $\mu$ and $\sigma$ denote the mean and
standard deviation of the Gaussian component, and $\tau$ is the mean of the
exponential component. The probability density function (PDF) of their sum
$X = G + E$ can be written as
\begin{equation}
	f(x) = \int_{-\infty}^{x}
	\frac{1}{\sqrt{2\pi}\sigma}
	\exp\!\left[-\frac{(u-\mu)^2}{2\sigma^2}\right]
	\frac{1}{\tau}\exp\!\left[-\frac{(x-u)}{\tau}\right] du .
	\label{eq:conv_exgauss}
\end{equation}

Rearranging the exponentials and extracting $x$ independent factors yields
\begin{equation}
	f(x)
	= \frac{e^{-x/\tau}}{\tau\sqrt{2\pi}\sigma}
	\int_{-\infty}^{x}
	\exp\!\left(
	-\frac{(u-\mu)^2}{2\sigma^2} + \frac{u}{\tau}
	\right) du .
	\label{eq:exgauss_step1}
\end{equation}
The exponent in the integrand can be rewritten by completing the square in $u$:
\begin{equation}
	-\frac{(u-\mu)^2}{2\sigma^2} + \frac{u}{\tau}
	= -\frac{1}{2\sigma^2}
	\left[u - \left(\mu + \frac{\sigma^2}{\tau}\right)\right]^2
	+ \frac{\mu}{\tau} + \frac{\sigma^2}{2\tau^2}.
	\label{eq:exgauss_square}
\end{equation}
Substituting Eq.~(\ref{eq:exgauss_square}) into Eq.~(\ref{eq:exgauss_step1}) gives
\begin{widetext}
\begin{equation}
	f(x)
	= \frac{1}{\tau\sqrt{2\pi}\sigma}
	\exp\!\left(\frac{\mu}{\tau}+\frac{\sigma^2}{2\tau^2}-\frac{x}{\tau}\right)
	\int_{-\infty}^{x}
	\exp\!\left[-\frac{(u-m)^2}{2\sigma^2}\right] du ,
\end{equation}
\end{widetext}
where $m = \mu + \sigma^2/\tau$.
The remaining integral corresponds to a truncated Gaussian, which can be evaluated using the standard normal cumulative distribution function $\Phi$:
\begin{equation}
	\int_{-\infty}^{x}
	\exp\!\left[-\frac{(u-m)^2}{2\sigma^2}\right] du
	= \sigma\sqrt{2\pi}\;
	\Phi\!\left(\frac{x-m}{\sigma}\right).
\end{equation}
Inserting this result yields the compact analytic form of the ex-Gaussian PDF:
\begin{equation}
	f(x)
	= \frac{1}{\tau}
	\exp\!\left(\frac{\mu}{\tau}+\frac{\sigma^2}{2\tau^2}-\frac{x}{\tau}\right)
	\Phi\!\left(\frac{x-\mu}{\sigma}-\frac{\sigma}{\tau}\right),
	\label{eq:exgauss_phi}
\end{equation}
where $\Phi(z)$ is the cumulative distribution function of the standard normal distribution.

Using the relation
$\Phi(z) = \tfrac{1}{2}\,\mathrm{erfc}(-z/\sqrt{2})$,
Eq.~(\ref{eq:exgauss_phi}) can also be expressed in terms of the complementary error function:
\begin{equation}
	f(x)
	= \frac{1}{2\tau}
	\exp\!\left(
	\frac{\sigma^2}{2\tau^2} - \frac{x-\mu}{\tau}
	\right)
	\mathrm{erfc}\!\left(
	\frac{\mu + \sigma^2/\tau - x}{\sqrt{2}\,\sigma}
	\right).
	\label{eq:exgauss_erfc}
\end{equation}

The mean, variance, and skewness of this distribution follow directly from
the properties of the convolution:
\begin{align}
	\mathbb{E}[X] &= \mu + \tau, \\
	\mathrm{Var}[X] &= \sigma^2 + \tau^2, \\
	\gamma_1 &= \frac{2\tau^3}{(\sigma^2 + \tau^2)^{3/2}} .
\end{align}
In the limit $\tau \to 0$, the exponential component vanishes and
$f(x)$ reduces to a standard Gaussian function.
Equation~(\ref{eq:exgauss_erfc}) therefore provides a closed-form expression
for the convolution of a Gaussian core with an exponential tail,
which can effectively describe asymmetric line broadening observed in experimental spectra.




\bibliographystyle{apsrev4-2}
\bibliography{sn-article-KDAR-v2.7}

\end{document}